\documentclass[lettersize,journal]{IEEEtran}
\usepackage{amsmath,amsfonts}
\usepackage{algorithmic}
\usepackage{array}
\usepackage[caption=false,font=normalsize,labelfont=sf,textfont=sf]{subfig}
\usepackage{textcomp}
\usepackage{stfloats}
\usepackage{url}
\usepackage{verbatim}
\usepackage{graphicx}
\usepackage{hyperref}
\usepackage{pifont}

\def\BibTeX{{\rm B\kern-.05em{\sc i\kern-.025em b}\kern-.08em
		T\kern-.1667em\lower.7ex\hbox{E}\kern-.125emX}}
\usepackage{balance}
\begin{document}
\title{The Security Overview and Analysis of 3GPP 5G MAC CE}
	
\author{
	\IEEEauthorblockN{
		Yuanyuan Yang\IEEEauthorrefmark{1}, 
		Jin Cao\IEEEauthorrefmark{1}, 
		Ruhui Ma\IEEEauthorrefmark{1}, 
		Sheng Li\IEEEauthorrefmark{1},
		and Hui Li\IEEEauthorrefmark{1}} 
	
	\IEEEauthorblockA{\IEEEauthorrefmark{1}State Key Laboratory of Integrated Services Networks, School of Cyber Engineering, Xidian University, China}
}

\maketitle
	
	\begin{abstract}
		To more effectively control and allocate network resources, MAC CE has been introduced into the network protocol, which is a type of control signaling located in the MAC layer. Since MAC CE lacks encryption and integrity protection mechanisms provided by PDCP, the control signaling carried by MAC CE is vulnerable to interception or tampering by attackers during resource scheduling and allocation. Currently, the 3GPP has analyzed the security risks of Layer 1/Layer 2 Triggered Mobility (LTM), where handover signaling sent to the UE via MAC CE by the network can lead to privacy leaks and network attacks. However, in addition to LTM, there may be other potential security vulnerabilities in other protocol procedures. Therefore, this paper explores the security threats to MAC CE and the corresponding protection mechanisms. The research is expected to support the 3GPP's study of MAC CE and be integrated with the security research of lower-layer protocols, thereby enhancing the security and reliability of the entire communication system.
	\end{abstract}
	

	\section{Introduction}
	\IEEEPARstart{I}{n} the 5G NR protocol stack defined by 3GPP, the MAC layer serves as the core sublayer of Layer 2, positioned between the PHY and RLC layers. The MAC entity is responsible for mapping logical channels to transport channels, with the mapping process being completed based on the multiplexing mechanism configured by Radio Resource Control (RRC) \cite{3gpp.38.321}. The MAC PDU (Protocol Data Unit), serving as the core data bearing unit of the MAC layer, consists of a series of bytes arranged in sequence. Among these data units, the MAC Control Element (CE) is a fundamental component of the MAC PDU. The position of the MAC CE within the MAC PDU follows the priority of the Logical Channel ID (LCID), where high-priority CEs are placed immediately after the MAC subheader for efficient parsing \cite{yi2012}. Unlike the semi static parameters configured via the RRC layer, the MAC CE provides embedded signaling for low latency and high reliability control interactions.
	
	The typical application scenarios of MAC CE mainly include: (1) Timing Advance (TA) adjustment, which aims to compensate for the delay in the wireless signal propagation process and ensure the timing synchronization of uplink transmission in communication systems; (2) Buffer Status Report (BSR), which exists in three forms: short BSR, long BSR, and truncated BSR. The base station can use them to obtain detailed information about the uplink data buffered in each Logical Channel Group (LCG), and then achieve reasonable uplink resource scheduling based on Quality of Service (QoS) awareness; (3)Power Headroom Reports (PHRs), which can help optimize uplink power control; (4) DRX command MAC CE, which can effectively balance the power consumption of the terminal device while meeting the service continuity requirements of the terminal device \cite{3gpp.38.321}.
	
	During the evolution of wireless communication technologies from LTE to 5G and 5G-Advanced, the standardized functionality of MAC CEs has continuously expanded. LTE Release 8 initially defined basic MAC CEs, such as Timing Advance Command, DRX Command, Power Headroom Report, and C-RNTI. With the introduction of 5G NR and subsequent enhancements, the design of MAC CEs has gradually met the demands of multi-dimensional scenarios. For instance, Release 15 introduced the Downlink TCI State Indication, Release 16 added the Uplink Preemption Indication (UL PI), and Release 17 brought the Timing Advance Report MAC CE for Non-Terrestrial Network (NTN) propagation delay compensation. Release 18 focuses on MAC CE extensions for mobility management and multi-connectivity scenarios, such as the Uplink Power Headroom Report, Transmission Configuration MAC CEs supporting coordinated operations of multiple TRPs/multi-panel, and optimized Secondary Cell Group Activation/Deactivation MAC CEs. Currently, the 3GPP TS 38.321 V18.5.0 specification has defined over 50 types of MAC CEs, covering core functions such as dynamic resource allocation, uplink and downlink synchronization, and security activation. MAC CEs have evolved from basic control signaling to core protocol components supporting multi-scenario, high-dynamic, and high-security requirements, providing flexible and efficient resource management capabilities for 5G-Advanced and future wireless networks \cite{lei2021}. 
	
	While 5G has robust security mechanisms for high-layer control plane signaling, including encryption and integrity protection, it lacks protection for critical fields within MAC CEs, which are the bearers of low-layer control signaling. This security deficiency allows attackers to intercept and tamper with control signaling via malicious base stations or over-the-air eavesdropping \cite{kim2019}\cite{liu2025}. For instance, tampering with TA values can cause terminal uplink desynchronization and communication termination; forging BSRs may mislead base stations into incorrect resource over-allocation, resulting in network congestion and resource waste; forging DRX commands can force terminals into prolonged sleep states, rendering services inaccessible. Persistent attacks can deplete base station resources, interrupt service links, and degrade network performance. Studies have shown that attacks targeting MAC CEs have a far greater system level impact than single user data leakage \cite{yu2019}\cite{shaik2015}\cite{rupprecht2019}.
	
	Moreover, when combined with other vulnerabilities, the vulnerabilities of MAC CEs can lead to more severe risks. Attackers can exploit protocol vulnerabilities to launch man-in-the-middle attacks and fake base station attacks, disrupt the synchronization of the hybrid automatic repeat request (HARQ) process, and deceive a large number of identities during the random access process to launch DDoS attacks on base station resources. In mobile edge computing scenarios, these security risks can be amplified, triggering cascading events such as signaling storms and resource hijacking, which ultimately impair the service availability of the communication system and the robustness of the network \cite{hussain2019} \cite{3gpp.36.321}. Despite various user plane security solutions proposed in academia, research on MAC CE protection lags behind technology deployment and has not yet formed a systematic protection framework.

	This paper focuses on MAC CE, delving into its development history, current security risks, and future prospects, hoping to provide practical and valuable references and insights for the development of 3GPP standards. Specifically, the contributions of this paper are mainly reflected in the following three aspects: 
	\begin{itemize}
		\item{We provide a comprehensive introduction to MAC CE technology, include its basic functions and structural composition, and list several typical types of MAC CE, enabling readers to have a comprehensive and clear understanding of MAC CE technology.}
		\item{We thoroughly investigate the security risks faced by privacy sensitive fields in MAC CE. For instance, the combination of different fields may lead to a series of problems, including tampering and location privacy leakage. We conduct a detailed analysis of these issues.}
		\item{Finally, we conduct an discussion on the potential security protection research trends of MAC CE, aiming to point the way for subsequent related research.}
		
	\end{itemize}

	The remainder of this paper is organized as follows. Section II introduces the basic functions, composition structure, and several typical MAC CEs. Section III presents the security risks faced by privacy-sensitive fields in MAC CEs, including tampering and location privacy leakage caused by the combination of different fields (current location/long-term location), and finally provides our security protection suggestions. Section IV summarizes the content of this paper and presents future prospects.
		
	\section{Overview of MAC CE}
	\subsection{Functional Introduction}
	The MAC CE is a critical element in the MAC layer of LTE and 5G networks, responsible for conveying control information. It primarily handles resource allocation, state management, and protocol control between the User Equipment (UE) and the base station, characterized by its efficiency and flexibility. MAC CEs can be utilized to transmit various types of information or commands, such as those related to beam change, DRX configuration, Power Headroom Report (PHR), BSR, or recommended bit rate.
	
	MAC CEs are transmitted within TBs. These TBs are carried over shared channels, namely the Physical Downlink Shared Channel (PDSCH) in the downlink, the Physical Uplink Shared Channel (PUSCH) in the uplink, or the Physical Sidelink Shared Channel (PSSCH) for sidelink communication. As part of the MAC PDU, a MAC CE is multiplexed within the same TB alongside other MAC SDUs and other MAC CEs \cite{song2023}.
	
	The MAC PDU consists of a sequence of MAC subheaders and their corresponding MAC payloads. Each MAC subheader contains a LCID field. This field indicates the nature of the MAC payload, following the subheader and its associated logical channel type. For MAC CE messages, their corresponding LCID field uses predefined reserved values. For instance, when the LCID field in a MAC subheader is set to a specific reserved value, it indicates that the payload immediately following this subheader is a MAC CE message for reporting power headroom. Different MAC CE types correspond to different reserved LCID values .
	
	The receiving MAC entity parses the LCID field of each MAC subheader. Upon identifying that the LCID is a value reserved for MAC CEs, the receiver recognizes that the subsequent payload is a MAC CE. It then determines the specific type of MAC CE based on that particular LCID value, enabling appropriate processing. This mechanism allows the network to implement dynamic resource allocation, scheduling, and power control for the UE.
	
	\subsection{Composition Structure}
	A MAC PDU comprises one or multiple MAC sub-PDUs. Each MAC sub-PDU corresponds to one of the following four types \cite{ahmadi2019}:
	\begin{itemize}
		\item{A standalone MAC subheader (used for MAC padding);}
		\item{A MAC subheader followed by a MAC SDU;}
		\item{A MAC subheader followed by a MAC CE;}
		\item{A MAC subheader followed by MAC padding.}
	\end{itemize}
	
\begin{figure*}[htbp]
		\begin{center}
			\includegraphics[width=0.8\textwidth]{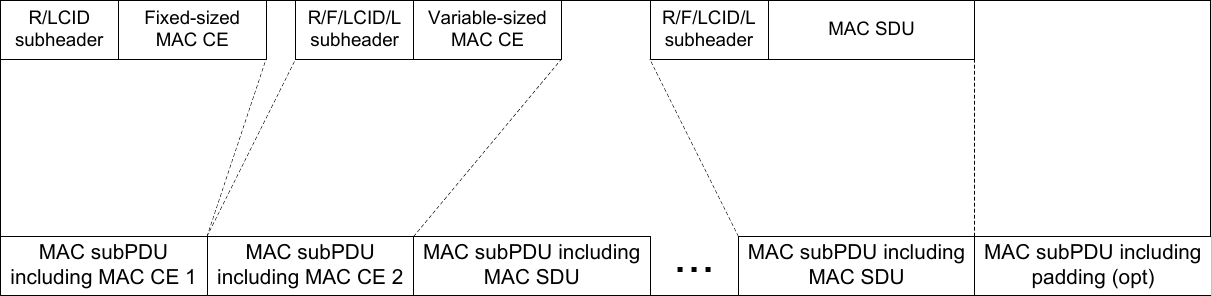}
		\end{center}
		\caption{Examples of Downlink MAC PDUs.}
		\label{fig_downlink}
	\end{figure*}

\begin{figure*}[htbp]
		\begin{center}
			\includegraphics[width=0.8\textwidth]{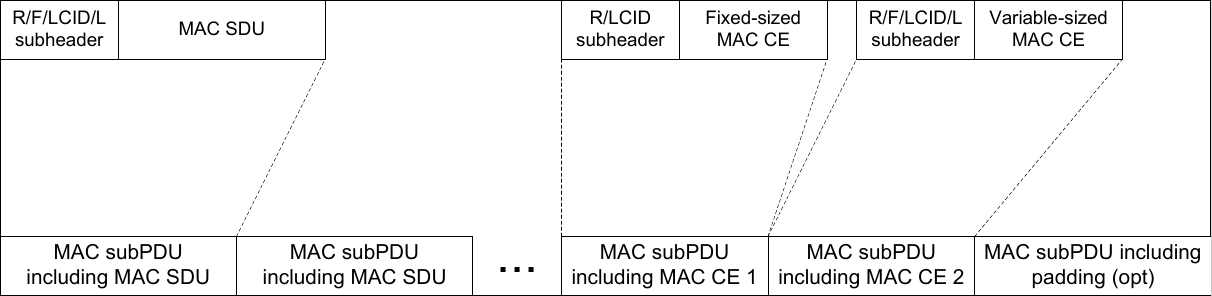}
		\end{center}
		\caption{Examples of Uplink MAC PDUs.}
		\label{fig_uplink}
	\end{figure*}

	In a Downlink (DL) MAC PDU (as illustrated in \ref{fig_downlink}, MAC CE sub-PDUs are aggregated within a contiguous region. This region is positioned before all MAC SDU sub-PDUs and precedes all padding sub-PDUs, ensuring priority processing of critical control information at the receiver. In an Uplink (UL) MAC PDU, as illustrated in \ref{fig_uplink}, MAC CE sub-PDUs are placed after all MAC SDU sub-PDUs and immediately before any padding sub-PDU. This structure guarantees that MAC CEs are not truncated by trailing padding.
	
	\subsection{Typical Types of MAC CEs}
	\begin{itemize}
		\item{\bf C-RNTI MAC CE.} The Cell Radio Network Temporary Identifier (C-RNTI) is a specific Radio Network Temporary Identifier (RNTI), which is dynamically allocated by the eNodeB/gNodeB to uniquely identify UE within a cell. During Radio Resource Control (RRC) connection establishment, the network assigns a C-RNTI to UE, which is subsequently embedded within MAC CE to convey critical MAC-layer control information, including scheduling grants, power control commands, and other directives essential for efficient wireless communication. The C-RNTI MAC CE structure, depicted in \ref{fig_CRNTI}, features a 16-bit C-RNTI field encapsulated following the MAC PDU subheader to enable high-priority transmission.

		\begin{figure}[htbp]
			\begin{center}
				\includegraphics[width=0.8\linewidth]{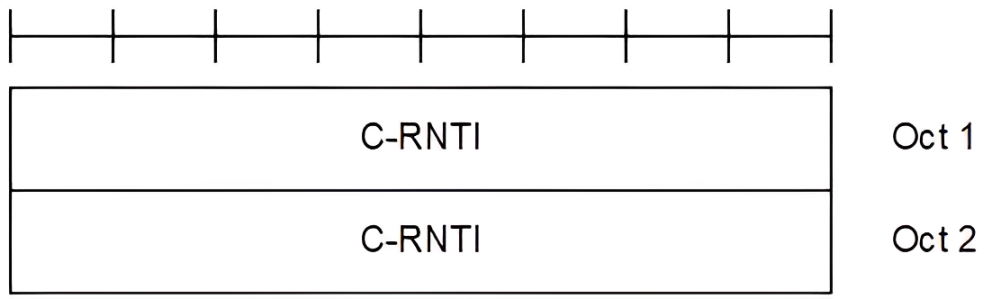}
			\end{center}
			\caption{Structure of C-RNTI MAC CE.}
			\label{fig_CRNTI}
		\end{figure}
		
		\item{\bf Timing Advance Report MAC CE.} It serves as a control signaling unit employed by UE to proactively report to gNB the TA calculated based on physical layer measurements. Its primary function is to assist the network in ensuring that the UE's uplink signals arrive precisely within the time window expected by the base station. This maintains uplink orthogonality and prevents inter-symbol interference and inter-user interference. The structure of this MAC CE is fixed at 2 bytes (16 bits) and is identified by a specific LCID in the MAC subheader. It comprises a 1-bit reserved field and a 14-bit TA value, as shown in \ref{fig_TA}. Key application scenarios for this MAC CE include accessing the target cell during handover, RRC re-establishment after radio link failure, and transitioning from inactive to connected states, where rapid uplink synchronization establishment or re-establishment is required. It is also crucial in high-speed mobility scenarios, such as those involving high-speed rail or drones, where dynamic TA adjustment is necessary to compensate for significant propagation delay variations caused by UE movement.
		\begin{figure}[htbp]
			\begin{center}
				\includegraphics[width=0.8\linewidth]{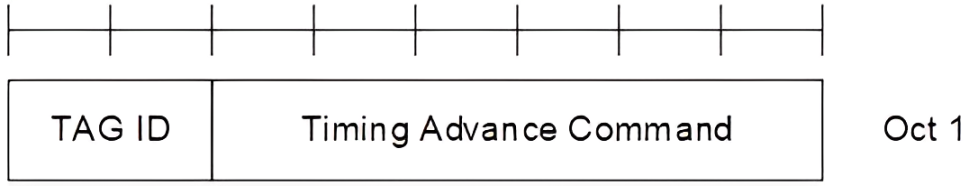}
			\end{center}
			\caption{Structure of TA Report MAC CE.}
			\label{fig_TA}
		\end{figure}
		
		\item{\bf SP CSI reporting on PUCCH Activation/Deactivation MAC CE.} The primary function of this MAC CE is to dynamically activate or deactivate the UE's Semi-Persistent (SP) CSI reporting configurations on the specified serving cell and uplink bandwidth part (UL BWP) via PUCCH. It contains a 5-bit serving cell ID, a 2-bit bandwidth part identifier (BWP ID), a 1-bit L field, and multiple Si fields (S0-Sn), as shown in \ref{fig_SP}. Each Si bit is mapped to a pre-configured SP CSI reporting configuration. By setting (1) or resetting (0) the bits, selective activation or deactivation of periodic CSI reporting tasks is achieved, with automatic triggering of status updates for associated sub-configurations. Compared to the traditional RRC reconfiguration mechanism, this MAC CE provides low-latency and highly flexible dynamic resource adaptation capabilities \cite{ludant2024}.
		
		\begin{figure}[htbp]
			\begin{center}
				\includegraphics[width=0.8\linewidth]{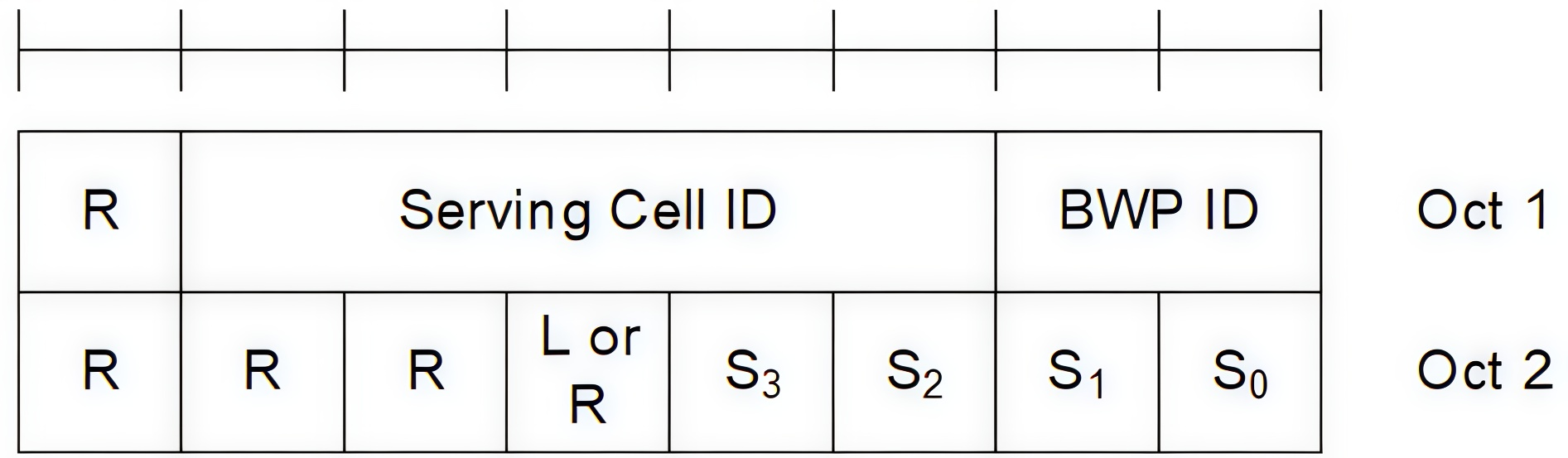}
			\end{center}
			\caption{Structure of SP CSI reporting on PUCCH Activation/Deactivation MAC CE.}
			\label{fig_SP}
		\end{figure}
		
		\item{\bf LTM Cell Switch Command MAC CE.} LTM is the core signaling mechanism for achieving low-latency mobility management in 5G NR.The LTM Cell Switch Command is triggered via a MAC CE, carries pre-configured parameters from the gNodeB to the UE. It includes target configuration ID, TA, and transmission configuration indication (TCI) state ID, as shown in \ref{fig_LTM}. To support inter-CU LTM, the LTM Cell Switch Command MAC CE may also include parameters such as the NCC, the selected algorithm indication for the target cell, and the security key set change indication. Compared to traditional RRC handover, the LTM mechanism significantly reduces signaling overhead and latency.
		\begin{figure}[htbp]
			\begin{center}
				\includegraphics[width=0.8\linewidth]{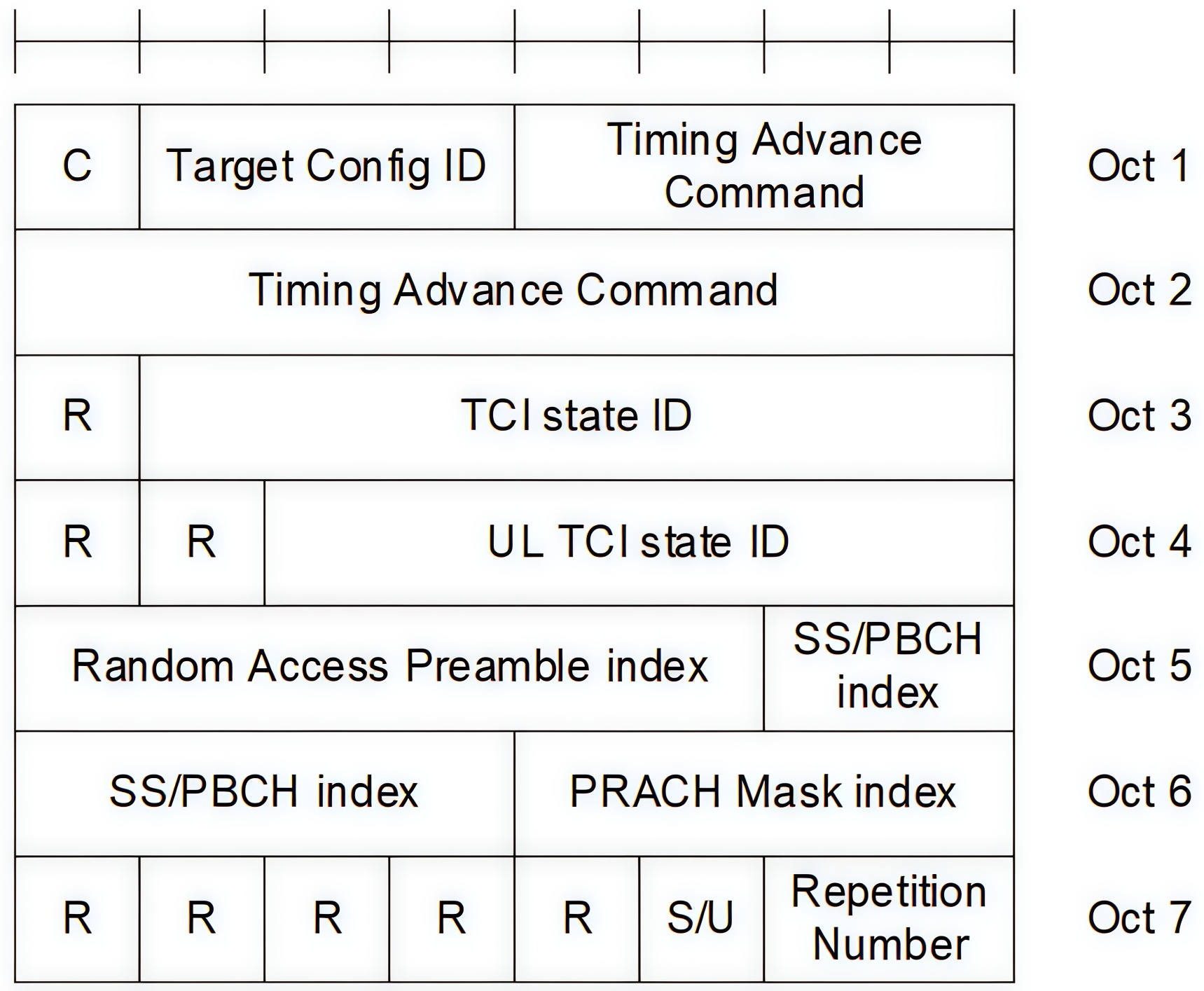}
			\end{center}
			\caption{Structure of LTM MAC CE.}
			\label{fig_LTM}
		\end{figure}
		
	\end{itemize}

	\section{The security risks of MAC CEs }
	\subsection{Analysis on Privacy-sensitive Field }
	The MAC CE serves as a pivotal mechanism for enabling dynamic radio resource allocation and adaptive beamforming optimization. However, the privacy risks associated with MAC CE fields haven't been adequately studied. Some fields may reveal sensitive user information to malicious entities. Some fields in the MAC CEs may cause privacy leakage due to carrying UE context or network parameters. Based on 3GPP specs, this section analyzes the privacy risks of the following MAC CE fields \cite{erni2022} \cite{kotuliak2022} \cite{yang2019} \cite{bitsikas2023} \cite{YU2021}:
	
	\begin{itemize}
		\item{\textbf{\emph{LCG ID:}}}
		The Logical Channel Group ID is used within MAC CEs to group multiple logical channels, supporting efficient radio resource management and scheduling operations. Different LCGs typically map to traffic flows or service types with distinct Quality of Service (QoS) requirements.
		
		\emph{Security Risk:} Different LCG IDs exhibit a strong correlation with different types of user traffic. An attacker passively monitoring LCG IDs within MAC PDUs can analyze the LCG ID usage patterns of specific UEs or user groups over time. Since different service types are typically configured to different LCGs or exhibit specific LCG usage patterns, the attacker can infer the application types currently or habitually used by the user, service preferences, and behavioral habits.
		
		\item{\textbf{\emph{C-RNTI:}}} The Cell Radio Network Temporary Identifier uniquely identifies a UE within a cell, used for downlink data transmission scheduling and uplink resource allocation.
		
		\emph{Security Risk:} As a long-term traceable temporary identifier, the C-RNTI enables an attacker to build a behavioral profile of a UE through passive monitoring. On the one hand, by long-term monitoring the signaling interaction patterns associated with a specific C-RNTI, attackers can establish associations between the C-RNTI and user behavior characteristics. On the other hand, leveraging publicly available base station geolocation databases, attackers can utilize the serving cell information of the C-RNTI to infer the user's frequented areas (e.g., residence, workplace).
		\item{\textbf{\emph{UE Contention Resolution Identity}}}
		Used during the random access procedure to resolve conflicts when multiple UEs simultaneously compete for radio resources, ensuring each UE is correctly identified and handled by the base station.
		
		\emph{Security Risk:} Although this identifier is temporary, it still faces security risks: Within the coverage area of a specific base station, an attacker monitoring the transmission of UE Contention Resolution Identities over the network can infer the user's active periods and daily network usage habits. Furthermore, if the attacker correlates contention resolution events across multiple base stations, they can exploit spatio-temporal correlations to deduce user movement paths.
		
		\item{\textbf{\emph{TAG ID:}}} This field identifies a Timing Advance Group (TAG). A TAG comprises a group of cells using the same uplink timing advance.
		
		\emph{Security Risk:} By monitoring changes in the TAG ID, particularly the addition, modification, or removal of non-zero TAGs, and correlating these with observed events like frequent handover measurement reports, an attacker can infer that the user is likely in a cell edge area or in motion. Further, if combined with significant and frequent fluctuations in the TA index value within the TA Command field, indicating continuous rapid adjustments to the uplink transmission timing, the inference that the user is in a high-mobility state can be strengthened, thereby leaking their mobility privacy.
		
		\item{\textbf{\emph{TA Command:}}} The base station transmits a TA command to guide the UE in adjusting the timing of its uplink transmission, thereby ensuring that the UE's uplink signals arrive at the base station on time and preventing temporal collisions among signals transmitted by different UEs.
		
		\emph{Security Risk:} The TA value is positively correlated with the UE-to-base-station distance. By long-term monitoring of the TA value variation patterns of a specific UE, an attacker can infer the user's behavioral patterns and frequented locations. For instance, if a relatively stable TA value associated with a specific geographical area is detected during nighttime, it implies that the user is currently at their residence; a different stable TA value during daytime suggests their workplace. This inference of the user's daily activity locations constitutes a severe privacy leakage risk.
		
		\item{\textbf{\emph{Serving Cell ID: }}} The serving cell ID uniquely identifies the cell that provides service to the UE, thereby facilitating the UE in conducting measurement reporting and enabling handover control \cite{ah2019}.
		
		\emph{Security Risk:} The Serving Cell ID is directly linked to the physical location of the cell. An attacker, leveraging publicly available base station latitude/longitude databases, can locate the UE within the cell's coverage area. Continuous monitoring of the Serving Cell ID allows the attacker to reconstruct the user's movement trajectory, leading to location privacy leakage.
		
		\item{\textbf{\emph{BWP ID:}}} Within a communication system, a carrier can be divided into multiple bandwidth parts; the Bandwidth Part IDentifier (BWP ID) identifies the currently activated bandwidth part for the UE.
		
		\emph{Security Risk:} BWP configuration is strongly correlated with traffic QoS. Typically, eMBB traffic activates large-bandwidth BWPs, mMTC traffic is allocated narrowband BWPs, and URLLC traffic enables low-latency BWPs. By analyzing the temporal variations of BWP IDs, the attacker can then reveal privacy information associated with the user's daily behavioral patterns.
		
		\item{\textbf{\emph{Spatial Relation Info ID:}}} It is a type of configuration information used to indicate that a signal (typically an uplink signal, such as SRS) should be aligned with or maintain a specific relationship with another known downlink reference signal in the spatial dimension.
		
		\emph{Security Risk:} The Spatial Relation Info ID has a deterministic mapping relationship with beam spatial parameters. By intercepting sequences of these IDs, attackers could infer the UE's direction using the base station beam pattern database, or calculate the user's position using triangulation. Moreover, attackers may infer the UE's movement speed and direction based on continuous changes in the Spatial Relation Info ID, resulting in location privacy leakage.
		
		\item{\textbf{\emph{SRS Resource's Cell ID: }}} The Sounding Reference Signal (SRS) resource specifies the serving cell that provides the physical resources in the SRS configuration signaling.
		
		\emph{Security Risk:} Attackers can leverage public cell location databases to map the SRS Resource's Cell ID to geographical coordinates, thereby locating users within the coverage area of the serving cell. If attackers are able to combine this ID with the value of TA, they may potentially elevate the positioning accuracy to the level of base station sectors. Additionally, attackers might reconstruct users' mobility tracks through sampling, thus leaking users' mobility behavior privacy.
		
		\item{\textbf{\emph{SRS Resource's BWP ID:}}} Indicates the bandwidth part identity allocated for the SRS resource, and its value maps to the physical layer Resource Block Group configuration.
		
		\emph{Security Risk:} Although this parameter itself is not sensitive, if attackers can monitor the user's bandwidth changes in real - time based on a BWP - service mapping knowledge base, they can infer specific service activities of the UE, such as whether it is playing high - definition video or downloading large files, leading to the leakage of user behavior patterns.
		
		\item{\textbf{\emph{Resource Serving Cell ID:}}} Identifies the physical cell that provides specific wireless resources (such as PDSCH/PUSCH). It is similar to the Serving Cell ID but emphasizes the serving cell related to specific resources.
		
		\emph{Security Risk:} Similar to the SRS Resource's Cell ID mentioned above, attackers can map the Resource Serving Cell ID to geographical coordinates via public global cell databases. If attackers can continuously monitor the changes in the Resource Serving Cell ID, they can reconstruct users' mobility tracks, thereby leaking users' location privacy.
		
		\item{\textbf{\emph{PCI:}}} The Physical Cell Identifier (PCI) Uniquely identifies a cell at the physical layer. During cell search, synchronization, and physical resource mapping processes, the UE identifies different cells by detecting the PCI.
		
		\emph{Security Risk:}The PCI broadcasted by 5G base stations is transmitted in plain text. Attackers may exploit this information to masquerade as legitimate base stations, attracting user devices to connect by amplifying the signal. Upon successful connection, attackers may acquire partial basic information of the mobile phone, such as the device model, and even disrupt normal user communication, for example, by disconnecting the user from the legitimate network and forcing them to connect to the rogue base station, thereby compromising communication security.
		
		\item{\textbf{\emph{TCI State ID: }}} It indicates the quasi-co-location (QCL) reference signal source configuration for the UE, which is used to determine the beamforming weights and antenna port mapping relationship.
		
		\emph{Security Risk:} Attackers may utilize the TCI state ID to reverse look up the SSB index of the serving beam. By combining this with publicly available base station location databases, they can potentially locate the UE . If attackers can obtain the historical TCI state IDs previously used by the UE and correlate these IDs with beam switching information, they may be able to track the user's mobility trajectory.
		
		\item{\textbf{\emph{SSB: }}}In 5G NR, this 6-bit index identifies specific SSBs and serves as a physical layer identifier for different base station beam directions, with each index corresponding to unique beamforming parameters \cite{palama2021}.
		
		\emph{Security Risk: } As the SSB Index is transmitted in the clear over the air, attackers can capture it in real-time with low cost sniffing devices. Attackers are capable of correlating the relevant information with the PCI to determine the specific location of the serving cell, and then estimating the physical location of the UE by utilizing the triangulation method. 
		
		\item{\textbf{\emph{Cell info ID:}}} It is an identifier utilized to specify the cell configuration associated with a specific resource set, which can be either a Child IAB-DU resource set or an IAB-MT resource set.
		
		\emph{Security Risk:} By collecting Cell Info IDs and their relationships with Child IAB-DU Resource Set IDs or IAB-MT Resource Set IDs, attackers can gradually build a topology map of the IAB network, understanding the connections and configurations of parent nodes, child nodes, and neighboring nodes. Moreover, attackers can infer the network's beam restriction or recommendation strategies, which reveals the network's wireless resource management policies.
		
		\item{\textbf{\emph{Candidate Cell ID:}}} During cell search, handover, and other processes of the UE, the Candidate Cell ID is used to identify the set of candidate cells, enabling the UE to perform further measurements and evaluations to determine whether to switch to these candidate cells \cite{3gpp.38.331}.
		
		\emph{Security Risk:} Its privacy risk lies in the Markov property of spatial topology. Attackers can infer the network topology around the UE based on the list of candidate cells with a public base station geographic information database. By continuously collecting candidate sets during working hours, attackers can infer the user's commuting route and map the user's daily behavior trajectory.
		
	\end{itemize}

	\subsection{Analysis on Tampering with Privacy-Sensitive Fields}
	
	Table~\ref{tab1} illustrates the potential security risks associated with the tampering of privacy-sensitive parameters by attackers. The number of stars in the third column signifies the magnitude of the security risk. A rating of five stars implies that if the parameter is tampered with, the security risk escalates to a catastrophic level. The security risk decreases progressively with the reduction in the number of stars. We have conducted an analysis of the security risks associated with the tampering of each parameter \cite{ludant2021}, as described below.
	\begin{itemize}
		\item{\textbf{\emph{C-RNTI: }}}When the user identity is hijacked, it allows the attacker to impersonate the user to receive data and causes legitimate users to be forcibly logged off.
		\item{\textbf{\emph{TA Command: }}}If the uplink TA is maliciously tampered with, the base station will be unable to properly demodulate user signals, thereby causing uplink synchronization failure. As a result, legitimate users will frequently experience disconnections.
		
		\item{\textbf{\emph{UE Contention Resolution Identifier: }}} When the identifier in the random access procedure is illegally altered, attackers gain the opportunity to seize user resources or trick users into transmitting sensitive information to a malicious base station.
		
		\item{\textbf{\emph{SSB Index: }}}Tampering with the SSB Index directly compromises the spatial alignment between the UE and the base station. This causes the uplink positioning SRS signal to be transmitted in an incorrect beam direction, leading to service connection interruption. Moreover, attackers may maliciously induce the UE to transmit signals in specific directions.
		
		\item{\textbf{\emph{Spatial Relation Info ID: }}} If the Spatial Relation Info ID is tampered with, the user signal may be misdirected, providing attackers with an opportunity to induce the user to transmit signals and subsequently obtain the user's location information.
		
		\item{\textbf{\emph{TCI State ID: }}} If the TCI State ID is modified, the user's beam may be directed towards the attacker or point towards an area with no signal coverage.
		
		\item{\textbf{\emph{PCI: }}}Tampering with the PCI can lead to users being induced to handover to rogue base stations or the attacker forging neighbor cell relationships, which can trigger frequent handovers by the UE.
		
		\item{\textbf{\emph{Cell Info: }}}If it is tampered with, sub-nodes may be misdirected to invalid or malicious network elements, or the transmission of critical beams may be unduly restricted. Once the backhaul link between the parent node and sub-nodes is interrupted due to configuration errors, all downstream nodes will be completely disconnected.
		
		\item{\textbf{\emph{TAG ID: }}} Improper manipulation of the TAG ID can cause signal interference among users within the same TAG and may also trigger communication conflicts in the uplink due to temporal overlap of multiple user signals.
		
		\item{\textbf{\emph{Resource Serving Cell ID: }}} If the resource serving cell ID is tampered with, users may incorrectly align their beams with unintended base stations, resulting in connection establishment failures.
		
		\item{\textbf{\emph{SRS Resource's Cell ID: }}}Tampering with the SRS resource's cell ID can cause the base station to obtain incorrect channel state information, leading to ineffective scheduling.
		
		\item{\textbf{\emph{LCG ID: }}}Tampering with the LCG ID can lead to high-priority services being mapped to low-priority channels and the UE being bandwidth-limited for specific services.
		
		\item{\textbf{\emph{Serving Cell ID: }}} If the Serving Cell ID is tampered with, users may incorrectly identify their serving cell, thereby triggering a signaling storm and ultimately leading the UE to request incorrect cell resources.
		
		\item{\textbf{\emph{Candidate Cell ID: }}} Once the Candidate Cell ID is tampered with, under the manipulation of an attacker, users may be induced to hand over to cells with excessive load or weak signals.
		
		\item{\textbf{\emph{BWP ID: }}}Tampering with the BWP ID can restrict users to a narrower bandwidth.
		
		\item{\textbf{\emph{SRS Resource's BWP ID: }}}Tampering with the SRS Resource's BWP ID can lead to incorrect channel measurements for the current BWP.

	\end{itemize}
	
	\begin{table*}[htbp]
		\begin{center}
			\caption{Risks from Privacy-Sensitive Parameter Disclosure.}
			\label{tab1}
			\renewcommand{\arraystretch}{1.5} 
			\begin{tabular}{| c | c | c | c|}
				\hline
				Privacy-sensitive parameters & The risk of tampering & Security risk rating \\
				\hline
				C-RNTI & Business Hijacking and	Denial of Service & \ding{73}\ding{73}\ding{73}\ding{73}\ding{73}\\
				\hline
				TA Command & Uplink Synchronization Loss& \ding{73}\ding{73}\ding{73}\ding{73}\ding{73}\\ 
				\hline
				UE Contention Resolution Identity & Access Hijacking and Rogue Base Station Inducement & \ding{73}\ding{73}\ding{73}\ding{73}\ding{73}\\
				\hline 
				SSB Index & Communication Interruption and Location Spoofing & \ding{73}\ding{73}\ding{73}\ding{73}\ding{73}\\
				\hline 
				Spatial Relation Info ID & Beam Misguidance and Location Exposure & \ding{73}\ding{73}\ding{73}\ding{73} \\
				\hline
				TCI State ID & Downlink Eavesdropping and Signal Jamming & \ding{73}\ding{73}\ding{73}\ding{73} \\
				\hline
				PCI & Handover Attack and Measurement Interference  & \ding{73}\ding{73}\ding{73}\ding{73} \\
				\hline
				Cell info & Incorrect beam configuration and backhaul link interruption & \ding{73}\ding{73}\ding{73}\ding{73} \\
				\hline
				TAG ID & Intra-Group Synchronization Disruption and Uplink Collision & \ding{73}\ding{73}\ding{73} \\
				\hline
				Resource Serving Cell ID & Inter-Base-Station Beam Misalignment & \ding{73}\ding{73}\ding{73} \\
				\hline
				SRS Resource's Cell ID & Channel Measurement Contamination & \ding{73}\ding{73}\ding{73} \\
				\hline
				LCG ID & QoS Degradation and Bandwidth Deprivation & \ding{73}\ding{73} \\
				\hline
				Serving Cell ID & Signaling Storm and Resource Allocation Error & \ding{73}\ding{73} \\
				\hline
				Candidate Cell ID & Erroneous Handover & \ding{73}\ding{73} \\
				\hline
				BWP ID & Reduced Transmission Rate & \ding{73} \\
				\hline
				SRS Resource's BWP ID & Localized Measurement Failure & \ding{73} \\
				\hline
				
			\end{tabular}
		\end{center}
	\end{table*}
	
	\subsection{Analysis on Location Exposure via Privacy-Sensitive Fields}
	In data privacy protection, the risk from isolated leakage of a single privacy-sensitive field is typically controllable. However, it is critical to focus on privacy threats that may emerge from the joint leakage of multi-dimensional data. We now analyze the privacy risks associated with the leakage of the UE’s current location privacy and long-term location privacy, respectively.
	
	\subsubsection{\textbf{UE Current Location Disclosure}}
	\begin{itemize}
		\item{\textbf{\emph{C-RNTI + Serving Cell ID + TA Command + SSB Index}}}
		
		Firstly, the C-RNTI serves as a unique identifier for the UE within the network. Secondly, by leveraging publicly available network databases, attackers may obtain geographical coordinate information that matches the Serving Cell ID. Subsequently, attackers utilize the TA value to calculate the distance between the UE and the base station, thereby narrowing down the potential location of the UE to an annular region centered on the base station. Finally, by incorporating the SSB Index, attackers further refine the UE's location to a specific sector within a certain azimuth angle. Moreover, if attackers are able to continuously track and record the historical TA values and SSB index sequences associated with the C-RNTI, they can analyze the UE's movement direction and speed, and consequently reconstruct the UE's movement trajectory.
		
		\item{\textbf{\emph{C-RNTI + PCI + TA Command + TCI State ID/Spatial Relation Info ID }}}
		
		Similar to the above analysis, the attacker maps the PCI to the geographical location of the base station based on the open database. The spatial angle relationship between the UE and the transmission reception point is determined by the TCI State ID/Spatial Relation Info ID. The attacker utilizes the TA value to calculate the relevant distance. By continuously monitoring the variations in both the TA and TCI State ID, reconstructs the motion vector of the UE.
		
		\item{\textbf{\emph{C-RNTI + SRS Resource's Cell ID +SRS Resource's BWP ID + Spatial Relation Info ID}}}
		
		The SRS Resource's Cell ID and BWP ID associated with the SRS resource serve to pinpoint the specific cell and bandwidth segment in which the SRS signal is being transmitted. The Spatial Relation Info ID indicates the beam used for SRS transmission. By detecting specific SRS resources and combining the Spatial Relation Info ID, Resource's Cell ID, and BWP ID, the attacker can accurately determine the uplink transmission beam direction of the UE. Ultimately, the UE's location can be pinpointed by integrating the base station's position and beamforming pattern.
		
	\end{itemize}
	
	\subsubsection{\textbf{UE Long-term Location Disclosure}}
	\begin{itemize}
		\item{\textbf{\emph{C-RNTI + Serving Cell ID / Resource Serving Cell ID + PCI}}}
		
		Initially, the attacker can correlate the new and old C-RNTI with the same UE by leveraging IMSI capturing or GUTI tracking. Subsequently, based on the Serving Cell ID, Resource Serving Cell ID, and PCI at different time intervals, the attacker can pinpoint the physical cell to which the UE is connected and the location where the resources are allocated. In this way, a map of the UE’s long-term activities can be delineated, thereby enabling the inference of the UE’s place of residence, workplace, and daily routine.
		
		\item{\textbf{\emph{C-RNTI + Serving Cell ID + Candidate Cell ID + Cell info ID}}}
		
		The Candidate Cell ID refers to the list of target cells configured by the network for the UE. The Cell Info ID (primarily used in IAB scenarios) implies the UE's network topological position within a multi-tier relay network. By continuously monitoring measurement reports or handover commands, the attacker can infer the UE's mobility trajectory.
		
		\item{\textbf{\emph{C-RNTI + SRS Resource's Cell ID / Resource Serving Cell ID + BWP ID + TAG ID +TA }}}
		
		The UE is allocated to distinct TAGs in accordance with its physical distance from the base station to achieve uplink synchronization. It follows that the leakage of the TAG ID can indirectly disclose the approximate distance range between the UE and the base station. When the network encompasses multiple cells/TRPs, the attacker can capitalize on the variations in TA values reported by different TRPs to conduct more accurate triangulation. In addition to this, by means of the C-RNTI and the Cell ID/Resource Serving Cell ID of the SRS resource, the attacker is able to identify the specific cell that a particular UE is currently utilizing for uplink sounding resources. To sum up, by leveraging this combination of methods, the attacker can track the UE’s location in real-time and construct a corresponding trajectory map.
		
	\end{itemize}
	
	\begin{table*}[htbp]
		\begin{center}
			\caption{Recommendations on Securing Privacy-Sensitive Parameters.}
			\label{tab2}
			\renewcommand{\arraystretch}{1.5} 
			\begin{tabular}{| c | c | c | c| c| c|}
				\hline
				Privacy-sensitive parameters & Confidentiality requirements & Integrity requirements & Latency-sensitivity & Recommended security mechanisms\\
				\hline
				LCG ID & \ding{73} & \ding{73} & \ding{73} & M1\\
				\hline
				
				BWP ID & \ding{73} & \ding{73} & \ding{73}\ding{73} & M1\\
				\hline
				
				SRS Resource's BWP ID & \ding{73} & \ding{73} & \ding{73}\ding{73} & M1\\
				\hline
				
				PCI & \ding{73}  & \ding{73}\ding{73}  & \ding{73}\ding{73}\ding{73}\ding{73} & M1\\
				\hline
				
				UE Contention Resolution Identity & \ding{73}\ding{73} & \ding{73}\ding{73}\ding{73} & \ding{73}\ding{73}\ding{73} & M2\\
				\hline 
				
				TAG ID & \ding{73} & \ding{73}\ding{73}\ding{73}  & \ding{73}\ding{73}\ding{73} & M2\\
				\hline
				
				TA Command & \ding{73} & \ding{73}\ding{73}\ding{73}\ding{73}& \ding{73}\ding{73}\ding{73}\ding{73} & M2\\ 
				\hline
				
				Serving Cell ID & \ding{73}\ding{73} & \ding{73}\ding{73}\ding{73}  & \ding{73}\ding{73} & M2\\
				\hline
				
				SSB Index & \ding{73} & \ding{73}\ding{73}\ding{73}  & \ding{73}\ding{73}\ding{73} & M2\\
				\hline
				
				SRS Resource's Cell ID & \ding{73}\ding{73} & \ding{73}\ding{73}\ding{73}  & \ding{73}\ding{73} & M2\\
				\hline
				
				Resource Serving Cell ID & \ding{73}\ding{73} & \ding{73}\ding{73}\ding{73}  & \ding{73}\ding{73} & M2\\
				\hline
				
				Candidate Cell ID & \ding{73}\ding{73} & \ding{73}\ding{73}\ding{73}  & \ding{73}\ding{73}\ding{73} & M2\\
				\hline
				
				C-RNTI & \ding{73}\ding{73}\ding{73} & \ding{73}\ding{73}\ding{73}& \ding{73}\ding{73}\ding{73}& M4\\
				\hline
				
				Spatial Relation Info ID & \ding{73}\ding{73}\ding{73} & \ding{73}\ding{73}\ding{73}\ding{73}  & \ding{73}\ding{73}\ding{73} & M4\\
				\hline
				
				TCI State ID & \ding{73}\ding{73}\ding{73} & \ding{73}\ding{73}\ding{73}\ding{73}  & \ding{73}\ding{73}\ding{73} & M4\\
				\hline
				
				Cell info ID& \ding{73}\ding{73}\ding{73} & \ding{73}\ding{73}\ding{73}\ding{73} & \ding{73}\ding{73}\ding{73}& M4 \\
				\hline
			\end{tabular}
		\end{center}
	\end{table*}

	\subsection{Security Protection Recommendations}
	
	Based on the specific security threats analyzed for MAC CEs, a differentiated security framework is proposed, comprising four mechanisms: plaintext direct transmission, integrity protection, encrypted transmission, and hybrid security. These mechanisms are designed to address the security risks associated with different types of MAC CEs.

	\begin{enumerate}
		\item{\textbf{\emph{M1(Plaintext Direct Transmission).}}} This mechanism is applicable when the MAC CE types are associated with low security risks. It operates in an insecure transmission mode, where control signaling is transmitted directly in its raw data form over an open channel.
		
		\item{\textbf{\emph{M2(Integrity Protection)}}} This mechanism is suitable for MAC CE types with medium security risks. It integrates hash algorithms (e.g., SHA-256) and message authentication codes (e.g., HMAC). By appending a dynamic verification code to the plaintext data, it ensures the verifiability and tamper resistance of the data during transmission.
		
		\item{\textbf{\emph{M3(Encrypted Protection).}}} This mechanism is employed when MAC CE types are confronted with relatively high security risks. It leverages symmetric encryption algorithms (e.g., AES-256) to implement end-to-end encryption of the transmitted information. It is particularly suitable for scenarios that require confidentiality and are highly sensitive to latency/cost.
		
		\item{\textbf{\emph{M4(Hybrid Security).}}}  This mechanism is tailor-made for MAC CE types with the highest security risks, offering both encryption and integrity protection for MAC CEs.
	\end{enumerate}
	
	This classified and graded security mechanism can flexibly adjust security strategies according to the characteristics of different MAC CE types, ensuring that security requirements are met while optimizing resource utilization. For low risk MAC CEs, plaintext transmission reduces control signaling processing latency. For MAC CEs with high tamper risk, the integrity protection mechanism is used. For resource configuration or user identity information related MAC CEs, the encryption mechanism prevents privacy leakage. For key security and positioning service related MAC CEs, the hybrid security mechanism maximizes security.
	
	As shown in Table \ref{tab2}, Based on the risk level, we conduct a categorization and grading of the privacy-sensitive parameters in MAC CEs and provide corresponding protection recommendations.

	\section{Conclusion}
	\noindent Currently, the MAC CE acts as a vital enabler for resource scheduling and signaling transmission in communication systems. However, the absence of robust encryption and integrity protection mechanisms renders it susceptible to significant security vulnerabilities. This paper concentrates on the privacy sensitive fields within MAC CEs and systematically elucidates the security threats they confront. Attackers can exploit these fields by intercepting or tampering with them, thereby causing user location privacy leakage, DoS attacks, and even impersonation attacks. To mitigate these risks, this paper proposes a security protection framework for MAC CEs. The security threats and protection outcomes presented in this paper can actively facilitate the inclusion of MAC CE security enhancement mechanisms in the core security specifications of the 3GPP R20 standard. This will lay a secure and trustworthy foundation for the safe and efficient access of massive Internet-of-Things (IoT) devices and the privacy preserving high precision positioning services in 6G networks.
	
	\bibliography{mybibfile}
	
\end{document}